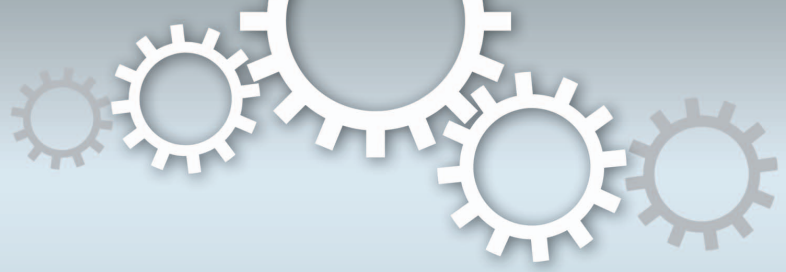

**OPEN**

SUBJECT AREAS:
HUMAN BEHAVIOUR
COOPERATION
EMPATHY
AGENCY

# Embodied social interaction constitutes social cognition in pairs of humans: A minimalist virtual reality experiment

Tom Froese[1,2,3], Hiroyuki Iizuka[4,5] & Takashi Ikegami[3]

[1]Departamento de Ciencias de la Computación, Instituto de Investigaciones en Matemáticas Aplicadas y en Sistemas, Universidad Nacional Autónoma de México, Mexico, [2]Centro de Ciencias de la Complejidad, Universidad Nacional Autónoma de México, Mexico, [3]Ikegami Laboratory, Department of General System Studies, University of Tokyo, Japan, [4]Human Information Engineering Laboratory, Department of Bioinformatic Engineering, University of Osaka, Japan, [5]Center for Information and Neural Networks, National Institute of Information and Communications Technology, Japan.

Scientists have traditionally limited the mechanisms of social cognition to one brain, but recent approaches claim that interaction also realizes cognitive work. Experiments under constrained virtual settings revealed that interaction dynamics implicitly guide social cognition. Here we show that embodied social interaction can be constitutive of agency detection and of experiencing another's presence. Pairs of participants moved their "avatars" along an invisible virtual line and could make haptic contact with three identical objects, two of which embodied the other's motions, but only one, the other's avatar, also embodied the other's contact sensor and thereby enabled responsive interaction. Co-regulated interactions were significantly correlated with identifications of the other's avatar and reports of the clearest awareness of the other's presence. These results challenge folk psychological notions about the boundaries of mind, but make sense from evolutionary and developmental perspectives: an extendible mind can offload cognitive work into its environment.

T he past couple of decades in the cognitive sciences have brought about profound changes in our understanding of the mind. Once mainly characterized in purely abstract computational terms of rule-based symbol manipulation, it is nowadays widely emphasized that our mind is embodied in a living organism[1] as well as extended into our concrete technological and social environment[2]. Perceptual experience is no longer seen as resulting from passive information processing, but as "enacted" via regulation of sensorimotor loops and active exploration of the environment[3–5]. The continuous changes taking place in our conscious experience, neural firing patterns, and sensorimotor interaction are beginning to be modelled and integrated in terms of dynamical systems theory[6–8]. The study of social cognition in more embodied and interactive settings has also been gaining in prominence[9–11], and a corresponding rise in "second-person" neuroscience, which studies the brains of two people in real-time interaction, has begun to reveal neural mechanisms that are specifically active during social encounters[12–14].

One of the major outstanding controversies in this area is whether dyadic interaction can sometimes play a *constitutive* role for an individual's social cognition, such that a part of social cognition is realized by social interaction itself[15]. Although it is widely recognized that social interaction is essential for children's development as well as for facilitating adult social cognition, many cognitive scientists prefer to view the mechanisms of cognition as being strictly limited to the brain of one individual[16–18]. This traditional individualist assumption has been challenged from a variety of disciplines, which emphasize that embodied interaction with the world and with other people can also do real cognitive work[19]. However, practical difficulties of studying complex human-to-human interactions in real-time have hampered progress in resolving this debate[20]. While there are good theoretical reasons for supporting an "extended" interpretation of social cognition[21], and agent-based models[22] as well as dynamical systems theory[23] have demonstrated the possibility of interactively enabled mind extension in principle, the theory of an extended-interactive social cognition is still lacking more solid experimental support.

One promising approach to coping with the complexities of human-to-human real-time interaction is to mediate the interaction process via a human-computer interface that reduces the scope of social expression to a more manageable level. This approach has the added advantage that it allows a full recording of the sensorimotor dynamics of the interaction, because the entire process unfolds in a virtual space. Several psychological studies using such highly constrained virtual environments have thereby succeeded in showing that the interaction process itself can sometimes spontaneously facilitate aspects of social cognition[24]. Using a minimalistic virtual reality interface that allows only linear movement in an invisible 1D space, and which provides on/off tactile feedback about the presence/absence of object contact, Auvray, Lenay and Stewart[25] have devised an interactive approach to social cognition research, which has come to be known as the "perceptual crossing paradigm" (Fig. 1). The experimental setup was inspired by classic psychological studies of children's sensitivity to social contingency, such as the "double video TV paradigm" pioneered by Trevarthen and colleagues[26–28]. Auvray et al. asked participants to locate the other active player's avatar while avoiding two nonresponsive distractor objects of equal size; one was static, the other object, called the "mobile lure" or the other's "shadow" object, exactly copied the partner's movements at a fixed distance – like an instantaneous replay condition (Fig. 2). The three objects that each participant could encounter were therefore only distinguishable by their differing *affordances for interaction*. Since only two of the three objects were mobile and only one of these was able to respond to contact, each object could potentially be identified by its unique affordances. For example, only a static object enables the possibility of relocation at always exactly the same point without any surprises, whereas only a moving object enables the possibility of spatiotemporally dispersed interactions such as chasing after it and losing track of it, while only a responsive object affords the possibility of mutual coordination of action, embodied communication, turn-taking, and imitation of each other's movement patterns. Participants were asked to click whenever they judged that they were currently interacting with the other's avatar.

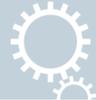

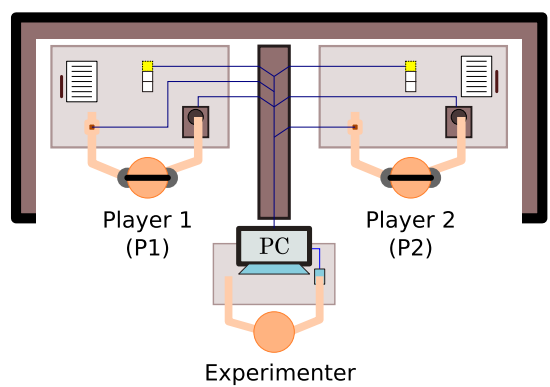

**Figure 1 | Experimental setup of perceptual crossing paradigm.** The two participants can only engage with each other via a human-computer interface that reduces their scope for bodily interaction to a bare minimum of translational movement and tactile sensation. Each player's interface consists of two parts: a trackball mouse that controls the displacement of their virtual "avatar", and a hand-held haptic feedback device that vibrates at constant frequency for as long as the avatar overlaps another virtual object and remains off otherwise. Three small lights on each desk signal the start, halftime (30 s), and completion of each 1-minute trial.

Auvray et al. found that participants were able to successfully locate each other in the virtual space, with most clicks being on target. However, surprisingly, participants seemed unable to consciously distinguish their partner's avatar from the moving distractor object: the probability of clicking after making contact with the partner was not significantly different from the probability of clicking after making contact with the other's shadow. Instead, the correct social judgments could be explained by the increased stability of mutual interaction. Since both participants actively look for each other, they tend to continue interacting when they happen to make mutual contact, while tending to move away from overly stable (a static object) and overly unstable (a non-responsive object) situations. The solution to the task, i.e. a participant's sensitivity to social contingency, was thus interactively realized.

Auvray et al.'s results have been confirmed by related experimental variations[24,29,30], and further supported by agent-based models[31–33]. However, a more radical interpretation of this experiment, namely that it has demonstrated that aspects of social cognition can be *constituted* by social interaction[15], has become the target of much criticism[34–36]. One of the caveats is that this interactive self-organization apparently leaves an individual's conscious or *explicit* social cognition about the other person, as measured by their clicks, unaffected: a regularly or randomly timed automatic click would have much the same objective outcome – since participants spend more time interacting with each other, clicks would tend to happen more often during their interaction. In other words, while the interactive self-organization of mutual localization is a nice example of implicit social scaffolding, this interactive process may have remained completely external to the minds of the participants and therefore falls short of conclusively demonstrating an interactive constitution of social cognition[34–36].

Accordingly, Lenay and Stewart[30] attempted to demonstrate that participants are also able to appropriate this interaction process from their first-person perspective. Arguing that participants need to be able to form an explicit memory trace of their past interactions, they introduced three different sound stimuli, one randomly assigned to each object, which were triggered upon contact. They asked players to identify the sound associated with the other's avatar after the end of a trial, rather than instructing participants to click whenever they judged to be interacting with the other player. The experiment was a success. For the first time the probabilities of object identification indicated an explicit recognition of the other's avatar, and it has been suggested that such external feedback is necessary to identify the source of stimulation[24]. However, this explicit appropriation of the interaction process by the individuals is still not necessarily an interactive achievement as such.

Importantly, recognition of a previously encountered object is enabled by the external sound signal alone, no matter whether the current interaction dynamics resemble those of a previous encounter. Thus, reliable identification of the other's avatar can simply be achieved by noticing which mobile sound is more frequently encountered. Although this experiment has demonstrated explicit social cognition, the introduction of interaction-independent markers for object discrimination has made it unclear whether this cognition is in fact interactively constituted. The new experiment thereby leaves the original critique unaddressed and the larger theoretical debate unresolved[37].

On our view, if a memory trace is in fact necessary for an experience of the other player's presence, then this recognition must also be interactively enabled by the two players without external aid, for example by co-developing a communication system on the basis of previous interactions. Only in this manner can it be guaranteed that any reported experience of the other player will be constituted by their interaction, rather than by some interaction-independent element of the experimental setup.

We employed the perceptual crossing paradigm to assess the ability of 17 pairs of participants (N=34) to recognize each other's presence only by engaging in interaction (Fig. 1 and Fig. 2). We hypothesized that previous studies using this paradigm had failed to find evidence for the interactive constitution of explicit social cognition, because they had neglected to set a collaborative task that required players to actively co-regulate their interaction in the service of communication or joint action[38]. We therefore specifically instructed players to form a team and to help each other in finding each other, and we ran the experiment as a tournament game to further motivate pair-wise cooperation. A team gained a point for every click that correctly identified the other's avatar, while losing a point for every wrong click. Only one click per player per trial was allowed (additional clicks were discarded); absence of a click was not penalized. Players were not informed about the successes of their clicks during the experiment such that no externally supervised (and therefore non-interactive) reinforcement learning was possible. Experimenters were also blind to the outcomes; clicking accuracy and team scores were only calculated after the completion of the study. Each team was tested for 15 trials with randomized avatar starting positions; each trial lasted one minute (see Fig. 3 for time series of an exemplary trial).

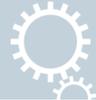

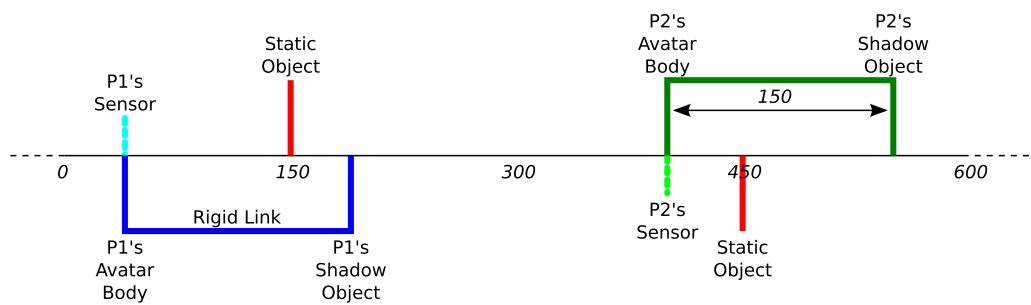

Figure 2 | **Virtual environment of perceptual crossing paradigm.** Players are virtually embodied as avatars on an invisible line that wraps around after 600 units of space. Each avatar consists of a binary contact sensor and a body object. Unbeknownst to the players a "shadow" object is attached to each avatar body at a fixed distance of 150 units. There are also two static objects, one for each player. All objects are 4 units long and can therefore only be distinguished interactively in terms of their different affordances for engagement.

## Results

The tournament was a success with most teams achieving high clicking accuracy and high scores (Table 1). After combining the data from all players, we find that median clicking accuracy was 92%. This is significantly better than chance level, and improves on the significant results found by related studies[25,29]. Accuracy was not achieved by a conservative strategy; a median of 10 clicks per player (out of a possible total of 15) correctly identified the other's avatar. We analysed the experiment from several perspectives.

**Objective behaviour: significance of social judgments.** The clicking accuracy could still fall short of demonstrating that social interaction constituted social cognition in this case, because in previous studies participants' high clicking accuracy was fully accounted for in terms of more frequent contacts with the other's avatar, rather than because of explicit social judgments[24]. We therefore determined the overall probability of a click following a single contact with a specific type of object. This calculation was based on 382 clicks from a total of 386 clicks; 4 clicks had no identifiable target. Contacts involving mutual sensory overlap were the most frequent type of encounter (28996 avatar contacts compared to 5046 shadow contacts and 5621 static object contacts), but this did not fully explain the proportion of correct clicks (338 avatar clicks compared to 33 shadow clicks and 11 static object clicks).

In contrast to previous studies, players' clicks were specifically sensitive to contact with the other's avatar: a click was nearly twice as likely to be made after a player was stimulated by the other's avatar (1.17% probability) when compared to the other's shadow object (0.65% probability), and almost six times as likely when compared to contact with a static object (0.20% probability). Judgments about the other's presence were thus specifically related to making contact with each other. These results confirm a prediction made on the basis of an agent-based model of the perceptual crossing paradigm, namely that turning the *individual epistemic* task of agency detection into a *social pragmatic* task aimed at mutual coordination would implicitly facilitate discrimination of the other's avatar[39].

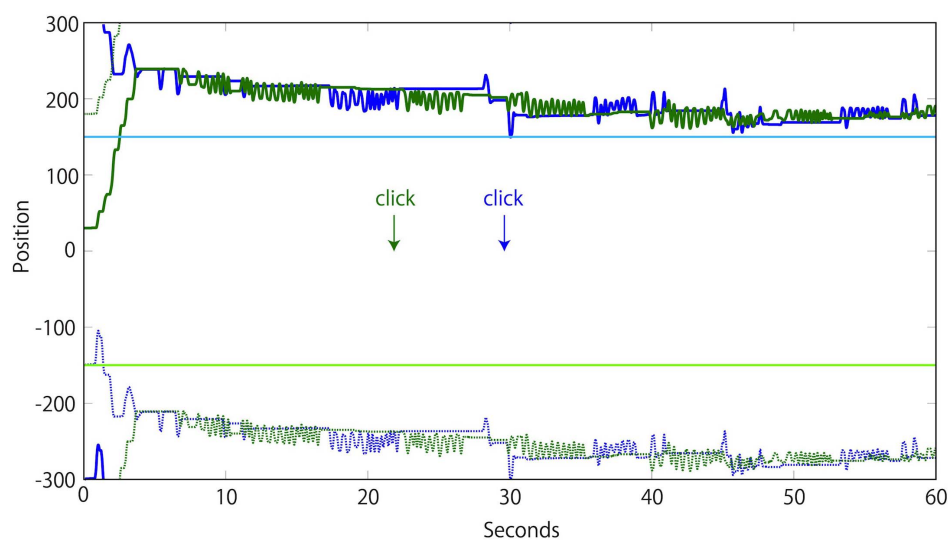

Figure 3 | **Time series of an illustrative trial.** Recording of the interaction between a pair of players across the 1D virtual environment (*y*-axis) over the 60-second duration (*x*-axis) of their 13th trial. Solid dark blue and dark green lines represent the positions of the two avatars, while the dotted blue and green lines represent the positions of their shadows. The solid light blue and light green lines represent the location of each player's specific static object. The arrows indicate the time of click for each player. From 5 s onwards a turn-taking (TT) interaction is noticeable (see the Methods section for a detailed explanation of how TT is measured). The green player's TT performance was nearly perfect because, during the 10 s before her click, there was a well-coordinated mutual exchange: first green stimulated while blue remained passive and then the roles were reversed (TT level of 0.88). The blue player's TT level was also good but slightly lower. There was a well-coordinated exchange of turns during much of the 10 s preceding her click, but this ended with a mismatch during the final seconds: blue failed to stimulate continuously in response to green's becoming passive (TT level of 0.52). Both players reported that they had a clear awareness of the other at the time of their click (PAS ratings of 4).

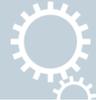

Table 1 | Results of the tournament game sorted by team score. The team score was calculated by summing both players' correct clicks (''Avatar clicks'') and subtracting both players' wrong clicks (all other clicks). The maximum achievable score was 30 (i.e. correct clicks by both players for all 15 trials). For each player the number of clicks per object type is shown, as well as those without any identifiable target object (''None'')

| Team score | Player 1: | | | | | Player 2: | | | | |
|---|---|---|---|---|---|---|---|---|---|---|
| | Avatar clicks | Shadow clicks | Static clicks | None clicks | Clicking accuracy | Avatar clicks | Shadow clicks | Static clicks | None clicks | Clicking accuracy |
| 27 | 12 | 0 | 0 | 0 | 100% | 15 | 0 | 0 | 0 | 100% |
| 26 | 13 | 0 | 0 | 0 | 100% | 13 | 0 | 0 | 0 | 100% |
| 24 | 12 | 1 | 0 | 0 | 92% | 13 | 0 | 0 | 0 | 100% |
| 23 | 12 | 0 | 0 | 0 | 100% | 11 | 0 | 0 | 0 | 100% |
| 22 | 13 | 0 | 0 | 0 | 100% | 9 | 0 | 0 | 0 | 100% |
| 22 | 10 | 0 | 0 | 0 | 100% | 12 | 0 | 0 | 0 | 100% |
| 19 | 9 | 1 | 0 | 0 | 90% | 11 | 0 | 0 | 0 | 100% |
| 19 | 14 | 0 | 0 | 0 | 100% | 8 | 2 | 1 | 0 | 73% |
| 19 | 11 | 1 | 0 | 0 | 92% | 10 | 1 | 0 | 0 | 91% |
| 18 | 8 | 0 | 0 | 0 | 100% | 12 | 1 | 1 | 0 | 86% |
| 18 | 11 | 0 | 0 | 0 | 100% | 10 | 3 | 0 | 0 | 77% |
| 16 | 11 | 1 | 0 | 0 | 92% | 8 | 1 | 1 | 0 | 80% |
| 12 | 10 | 2 | 0 | 1 | 77% | 6 | 0 | 0 | 1 | 86% |
| 10 | 9 | 0 | 1 | 0 | 90% | 6 | 1 | 2 | 1 | 60% |
| 10 | 9 | 1 | 1 | 0 | 82% | 3 | 0 | 0 | 0 | 100% |
| 6 | 9 | 2 | 1 | 0 | 75% | 6 | 3 | 2 | 1 | 50% |
| −1 | 5 | 7 | 1 | 0 | 38% | 7 | 5 | 0 | 0 | 58% |

Further analysis of clicking behaviour reveals that interactions between participants were highly coordinated in at least two respects. First, the success of clicks was highly co-dependent within trials: players were nearly twice as likely to jointly click correctly than for only one of them to click correctly alone. There were 131 trials of "Joint Success," during which both players clicked correctly, compared with 76 trials of "Single Success," during which only one player clicked successful and the other player clicked wrongly or not at all; there were 48 wrong clicks. Second, the timing of clicks in Joint Success trials was highly co-dependent even though participants could not directly sense the occurrence of each other's clicks. Clicks tended to occur close in time and the highest proportion happened in less than two seconds of each other. The near synchrony of recognition is due to the social interaction between the players; it cannot be explained in terms of independent entrainment to a shared time signal such as the duration of the trial (Fig. 4). This indicates that social judgments were not so much based on an individual recognition of *the other* but rather on a mutually shared recognition of *each other*, i.e. on an interactively shared cognitive process.

**Subjective ratings: Perceptual Awareness Scale (PAS).** In order to determine if participants also consciously experienced their interactive success we recorded their subjective reports. Free-text descriptions occasionally gave the impression that there were shared experiences of each other's presence (Table 2), but this effect was difficult to quantify. To better assess the subjective effects of mutual interaction we measured the felt clarity of the other's presence with an adapted Perceptual Awareness Scale (PAS)[40], which is a reliable direct measure of consciousness[41]. After each trial in which they clicked, players were asked to give a PAS rating between 1–4: "Please select a category to describe how clearly you experienced your partner at the time you clicked: 1) No experience, 2) Vague impression, 3) Almost clear experience, 4) Clear experience."

We emphasize that whereas a click reflects a social judgment about an objective state of affairs with respect to the other's current spatial position, a PAS rating is a categorical representation of the subjective experience of the other's presence. In agreement with phenomenological approaches to intersubjectivity, the adapted PAS can be interpreted as a direct measure of *empathy*, understood as the experience of a thematic encounter with a concrete other[42]. However, since researchers of social cognition typically use the notion of empathy in a more restricted sense, in particular to refer to the experience of sharing emotions[43], in the following we will describe PAS ratings in terms of awareness of the other's *presence* in order to avoid confusion.

We analysed a total of 384 PAS ratings from a total of 386 clicks (two ratings were absent). In general, players most frequently reported a clear awareness of the other's presence (PAS 4), which fits with the high levels of clicking success. PAS levels 4, 3, 2 and 1 were reported 143, 121, 101 and 19 times, respectively. More specifically, the clearest awareness of the other was most often reported for clicks occurring in Joint Success trials (Fig. 5). In order to determine whether there is a significant difference between the average PAS ratings reported for clicks occurring in Joint Success, Single Success and Wrong Click trials we applied a one-tailed, two-sample equal variance $t$-test. The equality of variances was verified using an $f$-test for each comparison. The average PAS rating was significantly higher for Joint Success clicks than for Single Success clicks (mean$_{joint}$=3.11 PAS, mean$_{single}$=2.75 PAS, one-tailed Student's $t$-test, $P$=0.001), and nearly significantly higher than for wrong clicks (mean$_{wrong}$=2.89 PAS, one-tailed Student's $t$-test, $P$=0.072). Although it might be expected that the other player's presence should be experienced significantly more clearly for clicks occurring in Single Success trials when compared to wrong clicks, this was not the case (one-tailed Student's $t$-test, $P$=0.180); both conditions were most frequently associated with medium levels of awareness. Experience of the other's presence was therefore not directly based on the other's objective presence, i.e. on the correctness of a click per se, but rather on the reciprocity of the interaction process with the other, as indicated specifically by Joint Success trials. A similar trend is observed in terms of Confidence Ratings.

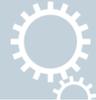
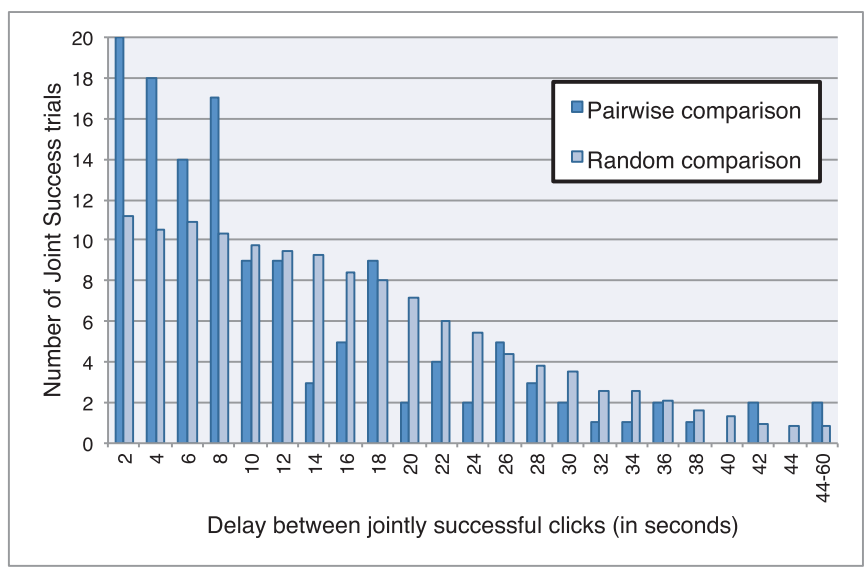

**Figure 4 | Jointly successful players synchronized with each other.** Clicks in Joint Success trials tended to happen close in time, even though players could not directly sense the occurrence of the other's click. The histogram shows the frequency of trials (y-axis) given a range of temporal delays between clicks (x-axis). The pairwise comparison represents the distribution of delays between two clicks *within* each Joint Success trial. The most frequent delay was for clicks to occur in less than two seconds of each other (20 trials). The random comparison represents the distribution of delays between randomly paired clicks *across* the set of Joint Success trials (we calculated the delays between all jointly successful clicks and normalized the results). Short delays (<10 s) tend to occur more frequently in the case of pairwise comparison than random comparison, which means that synchronization of players is an interactive achievement. It cannot be explained by their independent entrainment to an external factor (e.g. absolute trial time in seconds), since such a factor would be unaffected by the randomization of comparisons across trials.

**Subjective ratings: Confidence Ratings (CR).** We followed the suggestion by Auvray and Rohde[24] to measure Confidence Ratings (CR), which is a reliable indirect measure of conscious experience[44]. Post-trial questionnaires asked players to report a CR rating between 1–4 if they had clicked during the trial: "Please select a category to describe your confidence in your click's accuracy after the end of the trial: 1) No confidence, 2) Low confidence, 3) Medium confidence, 4) High confidence."

During the instructions at the start of the experiment it was strongly emphasized that whereas the PAS asks about the participant's experience at the time of the click, the CR scale asks about their confidence in the accuracy of the click after having completed the trial. The difference was highlighted to measure whether experience and confidence dissociated. For example, it is possible that a player clicked because of having felt a clear social experience at the time, but then later revised their confidence about the veridicality of this experience on the basis of subsequent interaction. The first three teams did not complete the CR questionnaire, so the results are only representative of the 14 subsequent teams.

We analysed a total of 305 CR ratings from a total of 321 clicks (16 ratings were absent). The distribution of CR ratings is qualitatively similar to that of PAS ratings: the highest level of confidence (CR 4) is most frequently reported overall, but especially for clicks in trials of Joint Success (Fig. 6). Medium levels of confidence (CR 3) were most frequently reported for correct clicks that were not matched by a partner's success (Single Success) and for wrong clicks. In other words, high confidence was not based on the objective co-location of the other player as such, but rather on mutual participation in the interaction (i.e. on Joint Success, but not Single Success, trials). However, there is also a slight dissociation between the two rating scales. In contrast to the PAS, there is a notable increase in frequency of the lowest CR category, i.e. absence of confidence (CR 1), from Joint Success to Single Success to Wrong Click trials.

Taken together these trends in CR ratings indicate that, by the end of a trial, players had a more accurate insight into the kind of situation in which they had been involved when they had clicked. Some players were able to realize on the basis of subsequent interaction that they had likely clicked in the absence of the other player's participation, as shown by the lower levels of confidence following Single Success and Wrong Click trials. This suggests an interaction-based possibility of learning, which could be a topic of future investigation.

**Objective ratings: turn-taking.** To objectively measure the amount of active co-regulation taking place during an interaction, we evaluated the extent of turn-taking (TT) that occurred before each click. Our measure yields a player's TT level as a number within the range [0–1], with 1 representing a perfect coordination of turns (an illustrative trial is shown in Fig. 3). We calculated how much the pairs of players traded periods of active stimulation and passive reception during 10 seconds preceding a player's click (see Methods for details). The results show that the level of TT performance was positively correlated with cooperation.

The comparison of average TT performance preceding clicks that occurred in Joint Success, Single Success and Wrong Click trials was done using the one tailed *t*-test. Since an *f*-test revealed unequal variances between TT levels in Joint Success and Wrong Click conditions, we used a two-sample unequal variance *t*-test for those comparisons. An *f*-test revealed equal variances for TT levels in Joint Success and Single Success, as well as Single Success and Wrong Click conditions, so a two-sample equal variance *t*-test was used to compare them. TT was significantly higher before clicks in Joint Success trials compared to Single Success trials (mean$_{joint}$=0.23 TT, mean$_{single}$=0.14 TT, one-tailed Student's *t*-test, $P=2.6\times10^{-4}$), and compared to wrong clicks (mean$_{wrong}$=0.11 TT, one-tailed Welch's *t*-test, $P=3.3\times10^{-7}$). Even though it might be expected that TT performance was significantly higher before correct clicks in Single Success trials when compared to wrong clicks, this was not the case (one-tailed Student's *t*-test, $P=0.15$). As with the subjective ratings, TT is therefore not an indication of objective clicking success per se, but rather of cooperative interaction.

TT performance was also positively correlated with a clearer awareness of the other's presence. The statistical comparison of average TT performance associated with each PAS level followed the same procedure. An $f$-test determined the variances to be equal, so a one-tailed, two-sample equal variance $t$-test was applied in each case. Average TT is significantly higher for PAS 4 compared to PAS 3 (mean$_{PAS\ 4}$=0.23 TT, mean$_{PAS\ 3}$=0.18 TT, one-tailed Student's $t$-test, $P$=0.007), and especially compared to PAS 2 (mean$_{PAS\ 2}$=0.15 TT, one-tailed Student's $t$-test, $P$=3.4×10$^{-5}$). There was no significant difference between PAS 4 and PAS 1 (mean$_{PAS\ 1}$=0.19 TT, one-tailed Student's $t$-test, $P$=0.16), but this may have resulted from an insufficiently representative sample of PAS 1 ratings (see next subsection for further discussion). These analyses confirm related findings showing that TT tends to be spontaneously co-developed in situations that require a sensitivity to social contingency, for example to create a novel embodied communication protocol[45] and for assessing each other's real-time responsiveness[46].

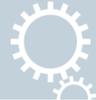

Table 2 | Example of trial-per-trial free-text subjective reports. Participants were asked to very briefly describe their experience of the other's presence after each trial. The following example illustrates a typical development across the 15 trials from the players' perspectives. In the first few trials the other's presence is obscured, but soon players start to become aware of the other's presence by noticing themselves to be the object of the other's searching movements. After a few more trials a turn-taking (TT) interaction starts to take shape, alongside a clearer presence of the other. This allows for some communicative intentions to be felt and shared. Some uncertainties arise while the conventions of mutual interaction are negotiated and modified over later trials, but co-presence tends to be re-established reliably. All clicks correctly identified the other player's avatar (PA).

| Trial | Player | Describe the sensation of your partner's presence: | Click | PAS | TT |
|---|---|---|---|---|---|
| 1 | 1 | I wasn't sure if it was actually the partner's avatar or just a moving object. | PA | 2 | 0 |
|   | 2 | She moved as if she is not just random, she seemed to know it's me. | PA | 4 | 0 |
| 2 | 1 | by the avatar's ''exploratory'' movement, | / | / | / |
|   | 2 |   | / | / | / |
| 3 | 1 | I think I ''found'' the partner's avatar quite close to the static object? So I wasn't sure if it was actually the avatar or the static object. | PA | 2 | 0.25 |
|   | 2 | vaguely | PA | 2 | 0.08 |
| 4 | 1 | I tried to compare the difference in sensation by looking for the static object and the partner's avatar. The difference was very clear. | PA | 4 | 0.48 |
|   | 2 | Ah, it's she! | PA | 4 | 0.43 |
| 5 | 1 | the partner's avatar was again moving around my avatar | PA | 4 | 0.22 |
|   | 2 |   | / | / | / |
| 6 | 1 | even when I stopped moving, I would receive a irregular feedback | PA | 3 | 0.17 |
|   | 2 | Clear experience! It's she! She is alive! | PA | 4 | 0 |
| 7 | 1 | I wasn't sure if it was a moving object or the partner's avatar. | PA | 3 | 0.20 |
|   | 2 | Ah, I come to understand now to recognize her! | PA | 4 | 0.24 |
| 8 | 1 | I wasn't sure if I got the partner's avatar. | / | / | / |
|   | 2 |   | / | / | / |
| 9 | 1 | I think there was a turn-taking-like communication. | PA | 4 | 0.52 |
|   | 2 | She likes me! | PA | 4 | 0.31 |
| 10 | 1 | It looks like we've established a way to communicate. | PA | 4 | 0.40 |
|   | 2 | I came to wonder if it is only an object. How can I feel it. | / | / | / |
| 11 | 1 | clear! | PA | 4 | 0.20 |
|   | 2 | Only objects cannot locate as this. | PA | 4 | 0.14 |
| 12 | 1 | presence of an intelligent object | PA | 4 | 0.40 |
|   | 2 | Clear, but afterward she escaped from me, which makes me wonder if that was really her, but only an object. | PA | 4 | 0.32 |
| 13 | 1 | wasn't sure if it was the partner's avatar | PA | 2 | 0.03 |
|   | 2 | Sorry I forgot clicking. | / | / | / |
| 14 | 1 | It felt clearly that we were trying to tell each other that we found ourselves. | PA | 4 | 0.30 |
|   | 2 | I guess it may be her, but she tends to escape (this was not clear to me) from me. | PA | 3 | 0.40 |
| 15 | 1 | Felt like she found me. | PA | 4 | 0 |
|   | 2 | It's she! | / | / | / |

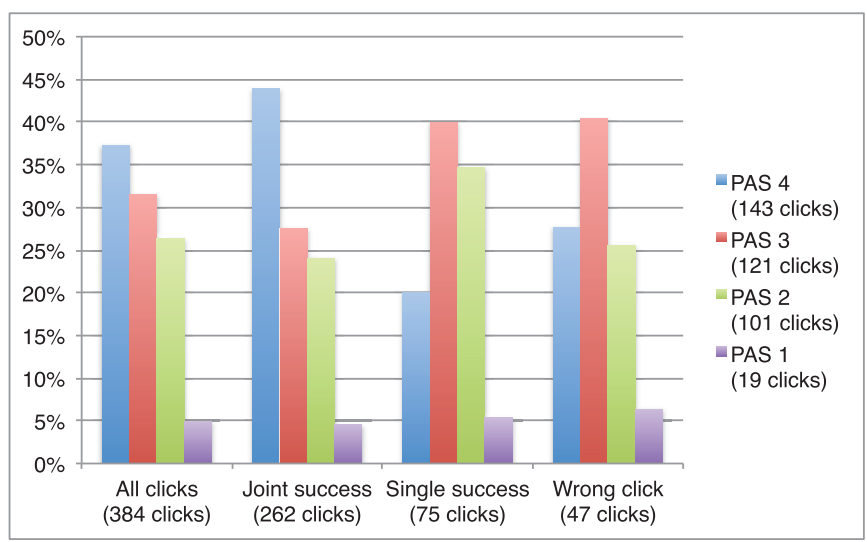

**Figure 5 | Relative frequency of Perceptual Awareness Scale (PAS) ratings.** The subcategory ''Joint Success'' includes ratings reported for clicks in trials during which both players clicked correctly; ''Single Success'' includes ratings reported for correct clicks in trials during which the other player clicked wrongly or not at all; ''Wrong Click'' includes ratings reported for wrong clicks regardless of the other's clicking success. The clearest experience of the other player's presence (PAS 4) was most frequently reported for clicks in Joint Success trials. This clarity subjectively distinguished cooperative situations; other clicks were most frequently associated with a medium level of social awareness (PAS 3).

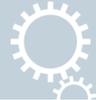

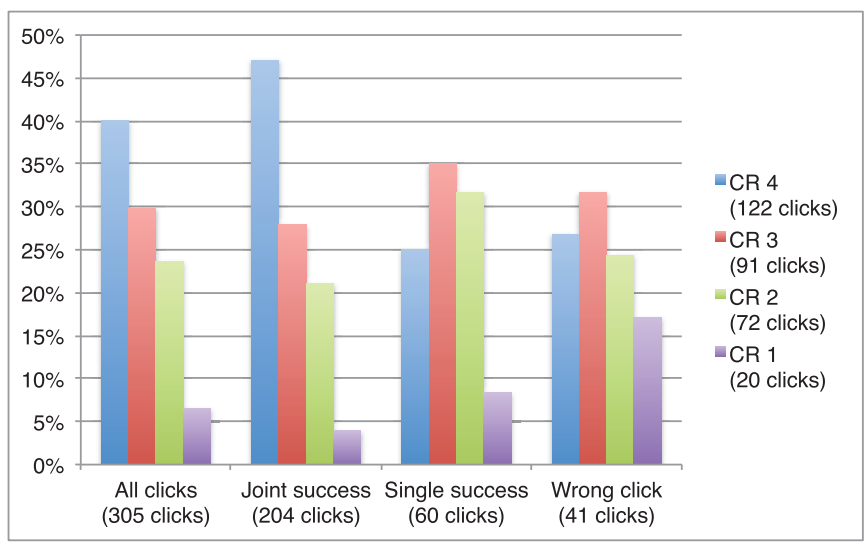

**Figure 6 | Relative frequency of Confidence Ratings (CR).** The subcategory "Joint Success" includes CR ratings reported for clicks in trials during which both players clicked correctly; "Single Success" includes ratings reported for correct clicks in trials during which the other player was unsuccessful (i.e. they clicked wrongly or not at all); "Wrong Click" includes ratings reported for wrong clicks (regardless of the other's success). The highest confidence in the accuracy of a click (CR 4) was most frequently reported after Joint Success trials. This confidence subjectively distinguished cooperative situations; other clicks were most frequently associated with a medium level of confidence (CR 3).

**Effects of a potential outlier.** We included all the data in our statistical analyses above, even though a potential outlier may partly explain the fact that (a) the PAS average for clicks occurring in Joint Success trials was not quite significantly higher than the PAS average for wrong clicks, and (b) the fact that the average TT level preceding clicks associated with PAS 4 reports was not significantly higher than the average TT level before clicks with PAS 1 reports. In particular, there was one highly skilled player who clicked 11 times, 100% correctly, out of which nine clicks occurred in Joint Success trials, and with a reasonable average TT performance of 0.3, but who consistently reported having had no experience of the other's presence for all clicks (PAS 1). When excluding that player from the statistics we end up with the expected trends: (a) the new PAS average for clicks occurring in Joint Success trials is significantly higher than the PAS average for wrong clicks (mean$_{joint}$=3.2 PAS, one-tailed Student's $t$-test, $P$=0.017), and (b) the average TT performance preceding PAS 4 reports is significantly higher than the new average TT performance before PAS 1 reports (mean$_{PAS\ 1}$=0.05 TT, one-tailed Student's $t$-test, $P$=0.001).

If we take this player's unusual PAS reports at face value, then it seems possible to be involved in a sufficiently reciprocal interaction to jointly achieve the task, yet without any conscious experience of the other's presence at all. This is compatible with a second-person approach to social cognition which hypothesizes that, in addition to interactional reciprocity, awareness of other minds also significantly depends upon emotional engagement[12], a factor which was not measured in the current experiment. The player may thus have employed a detached, non-engaged style of interaction. Future perceptual crossing experiments should include a measure of players' emotional engagement to determine whether this is in fact an essential factor in experiencing others. At the same time the discrepancy is in accordance with warnings from other studies of conscious experience, which have found a dissociation between subjective ratings and objective measures[47]. Our results confirm the methodological worry that direct measures of subjective experience cannot simply be replaced by objective measures of related behaviour[48].

## Discussion

Our hypothesis was that the perceptual crossing paradigm could be used to demonstrate that social interaction can in some cases constitute social cognition, without relying on elements that are independent from the interaction process itself. Following the enactive approach to perception, which holds that perceptual experience is constituted by active sensorimotor coordination[3,4], Froese and Di Paolo[38] proposed that participants in the original perceptual crossing study should be capable of a clear experience of the other, but only when they manage to mutually coordinate their sensorimotor interactions, for example in joint action. More precisely, "the task should be changed such that an intended activity of one participant can only become realized by the coordinated activity of the other"[38,23–24]. As an illustrative example we can think of the gesture of giving, which can only succeed if it is complemented by the gesture of receiving. The conditions of satisfaction of this kind of action go beyond individual agency[49]. We therefore replicated the original experimental setup, but with one crucial modification in the instructions: we told participants that they were engaged in a cooperative game and asked them to help each other in their task of identifying each other.

We made two predictions. We hypothesized on the basis of an agent-based model of the perceptual crossing paradigm[39] that those participants who managed to spontaneously develop a way of mutually coordinating their behaviour would exhibit more accurate clicking performance. And on the basis of the theoretical considerations of the enactive approach to perception and sociality[38], we hypothesized that those same participants would also clearly experience themselves as being engaged in social interaction, and thus demonstrate first-person awareness of the other's presence, specifically during that kind of mutual interaction.

Both of these hypotheses were confirmed. Most participants of the current study were able to interactively coordinate their embodied interactions in the minimal virtual space so as to create sufficient conditions for jointly becoming aware of each other's presence, and thus to click with higher accuracy. The fact that such co-regulation gave rise not only to a correct social *judgment* but also to an *experience* of the other's presence supports the enactive approach to social cognition. It gives empirical support to the theory that social interaction can sometimes partially constitute social cognition[15], especially when that interaction is co-regulated by participants[38]. It also backs the proposal that experiencing others is a perceptual modality constituted by exercising one's social skill of interacting with others[50]. As predicted by the enactive concept of "participatory sense-making"[51], we specifically found that mutual participation is constitutive of making sense of the social. These results challenge our folk psychological notions about the boundaries of mind[52], but make sense from evolutionary and developmental perspectives because an extendible mind can partially offload the mechanisms of cognition into its environment and thereby augment its capacities[23].

Going beyond cognitive extension, the results also have implications for the science of consciousness. The fact that a significant number of players simultaneously became aware of each other during their mutual dynamical entanglement is in accordance with the theory that the mechanisms of consciousness are also embodied in our comportment within the (social) world[53], and not just limited within our brain[54]. Whereas cognitive scientists have traditionally assumed that we are fundamentally isolated within our own heads, we suggest that we are actually open to genuinely sharing our minds with the other people around us – as long as we mutually participate in the unfolding of our embodied interaction.

## Methods

**Subjects.** Participants were healthy volunteers recruited from acquaintances at the University of Tokyo (N=34). There were 25 Japanese nationals, the rest were from various countries. Six were female. The mean age was 29 years. Teams were randomly created as volunteers became available. The Ethical Committee of the University of Tokyo approved the study. All of the participants gave their written informed consent before taking part in the study.

**Experimental setup.** At the start of an experiment participants were seated at two tables that were separated by a temporary wall (Fig. 1). On their desk they found a custom-made human-computer interface, active noise-cancelling headphones, a pen, and stapled papers consisting of the informed consent form, a detailed description of the experiment, and questionnaires. The interface consisted of three elements: 1) a commercially available trackball mouse, whose left- and rightward rotation changed the position of the player's avatar; 2) a small hand-held vibration motor, which would indicate any overlap between that avatar and another object by vibrating at constant speed, otherwise it would remain off; and 3) a series of 3 LEDs, which signalled the beginning of a trial (top light on), the middle of a trial after 30 seconds (middle light on), and the end of a trial (bottom light on). A change in light was also accompanied by a beep in the headphones to ensure that players were aware of the timing even when not looking at the LEDs.

**Task instructions and training procedures.** At the start of the experiment we followed Auvray et al.[25] in explaining the whole setup of the study and answering any questions participants may have, but without revealing the fact that the moving distractor object was actually rigidly fixed to the other's avatar at a distance. We explained that the moving object behaved similar to a player, but that it was merely following a pre-recorded trajectory and was therefore not responsive to contact. In contrast to other studies based on the perceptual crossing paradigm, we repeatedly emphasised that this was a cooperative game and that players formed a team. They were encouraged to work together to solve the task, but without specifying how they were to accomplish this cooperation in practice. Players were asked to devise a name for their team, which would later be used to anonymously rank their team score against the other teams participating in this tournament.

Again following Auvray et al., the players then familiarized themselves with the experimental equipment during a training phase, which involved interacting first with a static object, then with a slowly moving object and finally with a faster moving object, both moving at constant velocity. No interaction between the players was possible during this phase. Each of these objects was presented for the duration of a trial (1 minute) so that players could get a sense for how much time they had per trial. For the training with the static object players were encouraged to first locate the object and then to move away from it until they would encounter it once more. They were informed that this gave them a sense of the length of the 1D environment before it wrapped around. For the training with the moving objects players were asked to track the moving object as it moved around the circle as best as they could. Directly after training the proper experiment began.

**Testing procedures.** Players were asked to turn on the active noise cancellation of their headphones, and then to put them on; brown noise was played continuously. The experimenter started a trial as soon as both players held their human-computer interface in their hand. After each trial players were given a couple of minutes to complete the post-trial questionnaires and free text. After the end of the 15 trials they were given as much time as needed to provide additional free-text comments about their experience and general feedback about the experiment. Finally, players were debriefed about the purpose of the experiment, and the nature of the avatar-shadow link was revealed. The results of the tournament were e-mailed to each participant after the completion of all experiments.

**Analysis: identifying the target of a click.** Several choices had to be made with respect to the calculation of the intended targets of clicking behaviour, because it is often necessary to disambiguate situations in which a player is interacting with the other player's avatar or their shadow object in close proximity to the static object (there exists no spatial ambiguity between avatar and shadow because they are always set to be 150 units apart). We therefore chose a combined spatiotemporal measure: If the other's avatar (or shadow) is within range of a player's avatar one second before her click, assign the other's avatar (or shadow) as the intended target. The range $[-70, +70]$ was chosen so that the other's avatar and shadow cannot be within range at the same time. If neither the other's avatar nor their shadow were within range one second before her click, but the static object was within range, then we assigned the static object as the target. Otherwise, nothing was the intended target. This procedure assigned a cause to most clicks; only 4 out of 386 clicks had no discernable target.

**Analysis: identifying the source of a contact.** Several choices also had to be made about how to calculate the total number of contacts between an avatar and the three types of object, especially when there was overlap between two object types. We chose to define a player's "contact" as a string of uninterrupted tactile activations, with each object type contributing to this sequence being counted only once, even if involved in the same sequence at multiple distinct times. This choice reflects the player's first-person perspective on the situation. As an illustrative example we can consider the following fictitious time series (with "1" representing overlap between a player's avatar and another object, and "0" no overlap):

111000111100111000 (player's tactile stimuli)
-----------------------------------------------
100000111000000000 (other's avatar)
011000101100000000 (static object)
000000000000111000 (other's shadow)

The first tactile contact (three consecutive stimuli) consists of an overlap with the avatar lasting one time step, and another overlap with the static object lasting two time steps. The next tactile contact (four consecutive stimuli) consists of an overlap with the avatar lasting three time steps, and an overlap with the static object at two distinct moments lasting one time step and two time steps, respectively. The final contact (three tactile stimuli) consists of an uninterrupted overlap with the shadow object. Accordingly, both the avatar and the static object were involved in the first two tactile contacts, while only the shadow object was involved in the third contact, giving the following total number of contacts: avatar=2, static=2, and shadow=1.

**Analysis: identifying an exchange of turns.** Several decisions also had to be made regarding the calculation of a player's turn-taking (TT) performance. We wanted a measure that would exclude situations in which both players were continuously moving (or not moving) at the same time, or in which only one of the two players was continuously moving while the other was continuously stationary. At each time step we classified each player's behaviour in binary terms as either moving (1) or non-moving (0) by evaluating their computer mouse movement (we will refer to these behaviour time series as B1 and B2 for player 1 and player 2, respectively). Movement was considered to have taken place whenever the change of position $dx$ from one time step to the next was bigger than an $8^{th}$ of the avatar's body size (i.e. $4/8 = 0.5$ so that if $dx > 0.5$, 1, else 0). Since players often engaged in micro-saccades during their "turn" we chose to set a lower limit to the duration of pauses so as not to end up with microturns. Thus, we only set behaviours to 0 if there was no motion over at least 50 consecutive time steps (500 ms), otherwise they remain set to 1.

In order to determine a difference in activity, we applied the logical "Not-And" operator to these two time series (i.e. D = B1 Not-And B2). Then, we assigned to each player their active contribution of this movement exchange by applying the logical "And" operator and summing the result (i.e. C1 = sum(B1 And D); C2 = sum(B2 And D)). The overall TT performance for a given time period was then calculated by multiplying the players' active contributions and normalizing the outcome such that $TT = 4 * C1 * C2/T^2$, where $T$ is the number of time steps. The range of TT is therefore 0–1, with 0 representing a complete absence of TT interactions and 1 representing a perfect exchange of activity and passivity between the players. A trial with one of the highest TT levels is shown in Fig. 3. Data was recorded every 10 ms (100 Hz).

See Supplementary Information for further discussion of the experimental setup.

## Acknowledgments
We thank Marieke Rohde, Charles Lenay and Hanne De Jaegher for critical discussions and reading of the manuscript. Nele Froese helped with the statistics. T.F.'s running of this experiment at the Ikegami Laboratory was supported by a JSPS Postdoctoral Fellowship and a JSPS Grant-in-Aid. This work was also supported by a Grant-in-Aid for Scientific Research on Innovative Areas (#24120704 "The study on the neural dynamics for understanding communication in terms of complex hetero systems"). The funders had no role in study design, data collection and analysis, decision to publish, or preparation of the manuscript.

## Author contributions
T.F. developed the hypothesis of the study and wrote the paper; H.I. designed and constructed the haptic interface and programmed computer software; T.F. and H.I. recruited participants, supervised the study, and analysed the data; T.I. was involved in study design and data analysis. All authors discussed the results and reviewed the manuscript.

## Additional information
**Supplementary information** accompanies this paper at http://www.nature.com/scientificreports

**Competing financial interests:** The authors declare no competing financial interests.

**How to cite this article:** Froese, T., Iizuka, H. & Ikegami, T. Embodied social interaction constitutes social cognition in pairs of humans: A minimalist virtual reality experiment. *Sci. Rep.* **4**, 3672; DOI:10.1038/srep03672 (2014).